# A symmetry-based method to infer structural brain networks from probabilistic tractography data

Kamal Shadi[1], Saideh Bakhshi[2], David A. Gutman[3], Helen S. Mayberg[4], Constantine Dovrolis[5]

[1] PhD candidate,
School of Computer Science,
Georgia Institute of Technology,
Atlanta, GA, USA.
kshadi3@gatech.edu

[3] Assistant Professor,
Department of Neurology, Psychiatry & Biomedical Informatics
Emory University,
Atlanta, GA, USA.
dgutman@emory.edu

[2] Research scientist[#]
HCI Research Group,
Yahoo Labs,
San Francisco, CA, USA.
sbakhshi@yahoo-inc.com

[4] Dorothy C. Fuqua Chair in Psychiatric Neuroimaging and Therapeutics,
Professor of Psychiatry, Neurology, and Radiology,
Emory University,
Atlanta, GA, USA.
hmayber@emory.edu

[5] Professor,
School of Computer Science,
Georgia Institute of Technology,
Atlanta, GA, USA.
404-385-4205
constantine@gatech.edu
http://www.cc.gatech.edu/~dovrolis/

**Corresponding author:** Prof. Constantine Dovrolis (constantine@gatech.edu)

**Running head:** *Symmetry-based structural brain network inference*

**Keywords:** *Connectome, Diffusion MRI, Tractography, Structural Network, Network Analysis*

---

[#]Saeideh Bakhshi was a Ph.D student at Georgia Tech when she contributed to this work.

## Acronyms

| | |
|---|---|
| **AD** | Addictive Disorder |
| **DTI** | Diffusion Tensor Imaging |
| **FACT** | Fiber Assignment by Continous Tracking |
| **FDT** | FMRIB's Diffusion Toolbox |
| **fMRI** | functional Magnetic Resonance Imaging |
| **FMRIB** | Oxford centre for Functional Magnetic Resonance Imaging of the Brain |
| **fODF-PROBA** | Fiber Orientation Distribution Function-Probabilistic |
| **FSL** | FMRIB Software Library |
| **GAD** | Generalized Anxiety Disorder |
| **HCP** | Human Connectome Project |
| **MANIA** | Minimum Asymmetry Network Inference Algorithm |
| **MDD** | Major Depressive Disorder |
| **MNI** | Montreal Neurological Institute |
| **MRI** | Magnetic Resonance Imaging |
| **OCD** | Obsessive Compulsive Disorder |
| **PiCo** | Probabilistic Index of Connectivity |
| **PTSD** | Post-Traumatic Stress Disorder |
| **ROI** | Region Of Interest |
| **WGB** | White matter/Grey matter Boundary |

# Abstract

Recent progress in diffusion MRI and tractography algorithms as well as the launch of the Human Connectome Project (HCP)[1] have provided brain research with an abundance of structural connectivity data. In this work, we describe and evaluate a method that can infer the structural brain network that interconnects a given set of Regions of Interest (ROIs) from probabilistic tractography data. The proposed method, referred to as *Minimum Asymmetry Network Inference Algorithm (MANIA)*, does not determine the connectivity between two ROIs based on an arbitrary connectivity threshold. Instead, we exploit a basic limitation of the tractography process: the observed streamlines from a source to a target do not provide any information about the polarity of the underlying white matter, and so if there are some fibers connecting two voxels (or two ROIs) X and Y, tractography should be able in principle to follow this connection in both directions, from X to Y and from Y to X. We leverage this limitation to formulate the network inference process as an optimization problem that minimizes the (appropriately normalized) asymmetry of the observed network. We evaluate the proposed method on a noise model that randomly corrupts the observed connectivity of synthetic networks. As a case-study, we apply MANIA on diffusion MRI data from 28 healthy subjects to infer the structural network between 18 corticolimbic ROIs that are associated with various neuropsychiatric conditions including depression, anxiety and addiction.

---

[1] www.humanconnectome.org/

# 1 Introduction

Diffusion MRI has opened a new window at the meso-scale structure of the living brain (Sporns et al., 2005). Clinicians and researchers can now observe and measure the properties of white matter in a non-invasive manner, analyzing the location and density of neuronal fibers at a spatial granularity of 1-2mm isotropic voxels (Van Essen et al., 2013). Such structural information is important in deciphering how the brain works (Sporns, 2012, 2013), and it also creates new ways to understand and potentially diagnose (Ciccarelli et al., 2008; Fornito and Bullmore, 2015) or even treat (Mayberg et al., 2005) various brain diseases (Bassett et al., 2008; Buckner et al., 2005).

Processing diffusion MRI data using tractography algorithms is the next step forward: instead of analyzing the properties of white matter at the level of individual voxels, tractography aims to detect individual bundles of neuronal fibers originating or passing through a given "seed" voxel (Mori and van Zijl, 2002). Additionally, given a seed voxel and a target ROI, it is now possible to examine the likelihood that some white matter fibers connect the two (referred to as "probabilistic tractography"), and to track the shape of these connections (Behrens et al., 2007). In this paper, we propose a method to further process the noisy connectivity information provided by probabilistic tractography in order to estimate an interconnection network between a given set of grey matter ROIs.

Diffusion MRI data, jointly with deterministic (Mori et al., 1999) or probabilistic tractography methods (Behrens et al., 2007) have been used successfully during the last decade to infer the structure of the human brain between hundreds of ROIs (Hagmann et al., 2007). Various structural properties of these networks have been discovered for the healthy brain and for various psychiatric diseases (Daianu et al., 2013; McIntosh et al., 2008). When combined with fMRI and behavioral or genomic analysis, these non-trivial topological properties provide new insights about the role of individual ROIs in specific networks and the way in which these distinct ROIs exchange information to produce integrated function (Damoiseaux and Greicius, 2009; Greicius et al., 2009).

A major challenge in this research effort is that the inferred brain networks, as well as their topological properties, are often sensitive to the parameters of the tractography process (Bastiani et al., 2012; Duda et al., 2014; Rubinov and Sporns, 2010; Thomas et al., 2014). In probabilistic tractography, the most critical of those parameters is the *connectivity threshold* $\tau$ that determines whether the tractography-generated streamlines from a given seed voxel to a target ROI occur with sufficiently large probability to indicate the presence of an actual connection (L. Li et al., 2012b). If $\tau$ is too low the resulting network includes connections that do not exist in reality and the converse happens if $\tau$ is too high. Even a small number of spurious or miss-detected edges can adversely effect the properties of the inferred networks (Van Wijk et al., 2010). Further, the optimal value of $\tau$, i.e., the threshold that would result in the most accurate reconstruction of the underlying "ground truth" network, may vary between different subjects (Gong et al., 2009) and image acquisions parameters (Jones et al., 2013).

In the context of deterministic tractography, recent results question that the presence (or absence) of tractography-generated streamlines from a seed to a target is direct evidence that a connection exists (or that it is absent) (Jones et al., 2013; Reveley et al., 2015; Thomas et al., 2014). So, even in the case of deterministic tractography, it may be better to apply a threshold $\tau$ on the fraction of streamlines that appear to connect a seed voxel to the target, instead of taking these streamlines "at face value" as actual anatomic connections.

The problem of selecting an appropriate connectivity threshold in either deterministic or probabilistic tractography is not new. One approach has been to select the largest possible threshold (i.e., fewest possible edges) so that the final inferred network remains connected across the majority of subjects (Y. Li et al., 2009). Depending on the selected ROIs, this approach can lead to many miss-detections (if those ROIs are densely interconnected) or false alarms (if the ROIs are not directly connected). Another approach has been to analyze the tractography data with a wide range of threshold values, hoping that certain qualitative properties are robust and independent of the exact threshold. Li et al. investigated how the connectivity thresholds affects network density and therefore network efficiency metrics (L. Li et al., 2012b). Duda et al. have shown the importance of the connectivity threshold for various network metrics such as clustering coefficient and characteristic path length (Duda et al., 2014).

This work focuses on the following problem: *how to infer the structural network between a given set of grey matter ROIs in a reliable way that does not require an arbitrary choice of the connectivity threshold?* The proposed method, referred to as *Minimum Asymmetry Network Inference Algorithm (MANIA)*, exploits a fundamental limitation of diffusion MRI imaging and of the tractography process: diffusion MRI can estimate the orientation of fibers in each voxel but it cannot infer the polarity (afferent versus efferent) of those fibers (Jbabdi et al., 2011; Robinson et al., 2010). Similarly, a tractography algorithm can combine those per-voxel orientations to "stitch together" expected connections but it does not provide any information about the direction of those connections (Hagmann et al., 2008). Given this limitation, MANIA expects that the presence of an actual connection from voxel X to voxel Y (in that direction) will be detected by the tractography process as a symmetric connection between X and Y. Similarly, if there is no connection between X and Y, the tractography process should not detect a connection in either direction. Based on this principle, MANIA formulates the network inference problem as an optimization over the range of connectivity threshold values: it selects the value of $\tau$ that minimizes the asymmetry of the resulting network. The network asymmetry is normalized relative to the asymmetry that would be expected due to chance alone in a random network of the same density.

MANIA can work in tandem with all probabilistic tractography methods, such as FSL's probtrackx (Behrens et al., 2007), PiCo (Parker et al., 2003), and fDF-PROBA (Descoteaux et al., 2009). It can be also combined with deterministic tractography methods, such as FACT (Mori et al., 1999), but only if a large number of streamlines (in the thousands) are generated from randomly placed seeds within each voxel.

We expect that the given set of ROIs primarily reside in grey matter. Dilating a grey matter ROI so that it includes some white matter voxels may result in connectivity errors, especially with cortical ROIs, because of the dense white matter systems just beneath the cortical sheet (Reveley et al., 2015). The selection of ROIs and the estimation of their boundaries is an important issue that is further discussed in Section 4.

We evaluate the accuracy of MANIA based on synthetically generated data in which the ground-truth network is known. We also compare MANIA with an ideal threshold-based method in which the optimal connectivity threshold is assumed to be known. Further, we show how to associate a confidence level with each edge, and how to apply MANIA in a group of subjects. Finally, as a case-study, we apply MANIA on diffusion MRI data from 28 healthy subjects (Chen, Xu, et al., 2013) to infer the structural network between 18 corticolimbic ROIs that are implicated with neuropsychiatric disorders such as major depressive disorder (MDD), post traumatic stress disorder (PTSD), obsessive compulsive disorder (OCD), generalized anxiety disorder (GAD) and addictive disorder (AD) (Beucke et al., 2014; Elman et al., 2013; Peterson et al., 2014; Seminowicz et al., 2004). We note however that, even though these ROIs are generally associated with various psychiatric disorders and the aspects of emotional regulation putatively impacted by these disorders, the objective of this work is *not* to infer the network that is associated with any particular disorder.

## 2 Materials and Methods

### 2.1 DTI data, tractography parameters, and selected ROIs

We apply MANIA in Diffusion Tensor Imaging (DTI) data collected by an earlier study: *"Brain aging in humans, chimpanzees (Pan troglodytes), and rhesus macaques (Macaca mulatta): magnetic resonance imaging studies of macro- and microstructural changes"* (Chen, Xu, et al., 2013). Twenty eight healthy right-handed females between the ages of 18 and 22 (mean: 20.2), without a history of psychiatric disorder, were selected from that study. All subjects gave written informed consent, and the study was approved by the Emory University Institutional Review Board. Diffusion-weighted images were acquired using a Siemens 3T with a 12-channel parallel imaging phase-array coil. Foam cushions were used to minimize head motion. Diffusion MRI data were collected with a diffusion weighted SE-EPI sequence (Generalized Autocalibrating

Partially Parallel Acquisitions [GRAPPA] factor of 2). A dual spin-echo technique combined with bipolar gradients was used to minimize eddy-current effects. The parameters used for diffusion data acquisition were as follows: diffusion-weighting gradients applied in 60 directions with a b value of 1000 s/mm$^2$; TR/TE of 8500/95 ms; FOV of 216×256 mm$^2$; matrix size of 108×128; resolution of 2×2×2 mm$^3$; and 64 slices with no gap, covering the whole brain. Averages of 2 sets of diffusion-weighted images with phase-encoding directions of opposite polarity (left–right) were acquired to correct for susceptibility distortion. For each average of diffusion-weighted images, 4 images without diffusion weighting (b=0 s/mm$^2$) were also acquired with matching imaging parameters. The total diffusion MRI scan time was approximately 20 minutes. T1-weighted images were acquired with a 3D MPRAGE sequence (GRAPPA factor of 2) for all participants. The scan protocol, optimized at 3T, used a TR/TI/TE of 2600/900/3.02 ms, flip angle of 8°, volume of view of 224×256×176 mm$^3$, matrix of 224×256×176, and resolution of 1×1×1 mm$^3$. Total T1 scan time was approximately 4 minutes.

The resulting DTI data were processed using the FMRIB's Diffusion Toolbox (FDT) provided by FSL (FMRIB 4 Software Library) (Behrens et al., 2003). The FDT probabilistic tractography parameters were set to their default values (number of streamlines=5000, maximum number of steps=2000, loop check: set, curvature threshold=0.2, step length=0.5mm, no distance bias correction).

We applied MANIA on 18 corticolimbic ROIs. All ROIs are localized in Montreal Neurological Institute (MNI) standard space using the Automated Anatomical Labeling (AAL) (Jenkinson and Smith, 2001) of the WFU PickAtlas toolbox (Lancaster et al., 2000). The ROI acronym as well as the number of voxels in each ROI are shown in Table 1. The shape of these ROIs are not dilated and we adhere to the standard masks provided in the WFU PickAtlas toolbox. We chose these ROIs because they are known to play a significant role in various psychiatric disorders such as MDD, PTSD, OCD, anxiety and addiction (Craddock et al., 2009; James et al., 2009; Mayberg, 1997; Seminowicz et al., 2004). This sample of ROIs includes cortical, subcortical and limbic regions. Specifically, the cortical ROIs are BA6, BA9, BA10, BA40, BA46, BA47, the limbic are BA24, Th, BS, and the sub-cortical are BA11, BA25, BA32, Acb, Amg, Hp, Ht, Ins, Pc.

## 2.2 MANIA inputs

The proposed network inference method requires the following inputs:

1. A set of $N$ ROIs that represent the nodes of the structural brain network. The $i$'th ROI is a spatially connected cluster of $v_i$ voxels ($i = 1 \cdots N$). The selection of ROIs is important (de Reus et al., 2013b; Zalesky et al., 2010) but outside the scope of MANIA. MANIA attempts to find the anatomic network between the given ROIs independent of whether the latter are defined by an expert neuroanatomist or by a data-driven method. For instance, ROIs may correspond to different Brodmann areas or other anatomical atlases (Petrides, 2005; Tzourio et al., 1997). Or, it could be that the spatial extent of ROIs results from the analysis of fMRI data (Blumensath et al., 2013; Craddock et al., 2012; McKeown et al., 1997; Thirion et al., 2014). The results of tractography depend on the selection of ROIs, including their size, shape, grey/white matter composition, and their distance to other ROIs.

2. The results of the tractography process between the previous $N$ ROIs. We assume that the tractography results are structured as voxel-to-ROI matrices (i.e., streamlines are generated from each voxel towards each target ROI), instead of voxel-to-voxel or ROI-to-ROI matrices. Specifically, we represent the output of tractography with $N$ matrices $T_i$ ($i = 1 \cdots N$), defined as follows. The $i$'th ROI corresponds to a matrix $T_i$ with $v_i$ rows (i.e., the number of voxels in that ROI) and $N$ columns. The element $(j, k)$ of matrix $T_i$ represents *the fraction of tractography-generated streamlines that originate from the seed voxel $j$* of the $i$'th ROI *and reach any voxel of the $k$'th ROI*. The same number of streamlines is generated for every seed to target pair $(j, k)$. The $i$'th column of matrix $T_i$ is set to zero, meaning that we do not consider edges from an ROI back to itself (even if such fibers exist). Figure 1 illustrates this notation. Since we have $N$ ROIs, there will be $N$ input matrices, one for each source ROI.

### 2.3 Connectivity threshold $\tau$

How can we decide whether voxel $j$ of the $i$'th ROI connects to the $k$'th ROI given the fraction $T_i(j, k)$? The simplest approach is to examine if $T_i(j, k)$ is larger than a given *"connectivity threshold"* $\tau$ ($0 < \tau < 1$). In MANIA, $\tau$ is not a given threshold but an optimization variable, as described in the next section.

As in prior work, we assume that the $i$'th ROI is connected to the $k$'th ROI as long as at least one voxel of the former is connected to the latter (Bassett et al., 2011; Hagmann et al., 2008). This assumption is not central to MANIA however, and it can be easily replaced with a stronger connectivity constraint.

## 2.4 Network inference as an optimization problem

For a given value of τ, we can identify the voxels of the $i$'th ROI that connect to every other ROI. If this process is repeated for every $i$, we can construct a directed network in which the $i$'th ROI is connected to the $j$'th ROI if there is at least one voxel in the former that is connected to the latter for that value of τ. This network can be represented with an adjacency matrix[1] $G_\tau$, as follows:

$$G_\tau(i,k) = \begin{cases} 1 & if \ \ T_i(j,k) > \tau \ \ for \ at \ least \ one \ voxel \ j \\ 0 & otherwise \end{cases} \qquad (1)$$

So, the element $(i,k)$ of this matrix is equal to one if there is a (directed) edge from the $i$'th ROI to the $k$'th ROI. The diagonal entries of $G_\tau$ are set to zero because we do not consider streamlines from a source ROI back to itself.

We define the *asymmetry* $\phi(G)$ of a directed network $G$ as the fraction of edges that are present in only one direction,

$$\phi(G) = \frac{\sum_{i=1}^{N} \sum_{k=1}^{N} G(i,k)\big(1 - G(k,i)\big)}{\sum_{i=1}^{i=N} \sum_{k=1}^{N} G(i,k)} \qquad (2)$$

The asymmetry of a network $G$ depends on its *density* $\rho(G)$, defined as the fraction of connected node-pairs,

$$\rho(G) = \frac{\sum_{i=1}^{N} \sum_{k=1}^{N} G(i,k)}{N(N-1)} \qquad (3)$$

The more edges a directed network has, the more likely it becomes that a pair of nodes will be connected in both directions, i.e., the higher the density, the lower the asymmetry.

---

[1] In graph theory, an $N \times N$ adjacency matrix represents a directed and unweighted graph with $N$ nodes as follows: if there is an edge from node i to node j the (i,j) element of the adjacency matrix is 1; otherwise it is 0. The graph (and the corresponding adjacency matrix) are referred to as "weighted" if each edge is associated with a weight, which typically represents the strength of the edge.

More formally, consider a directed network with K directed edges and N nodes. The density is $\rho = K/[N(N-1)]$ $(0 < \rho < 1)$. To construct such a network randomly, denoted by $G_\rho$, we simply connect $N(N-1)\rho$ randomly selected but distinct pairs of nodes with directed edges. The expected number of edges that exist in only one direction is $N(N-1)\rho(1-\rho)$ and so the expected value of the asymmetry of $G_\rho$ is:

$$\bar{\phi}(G_\rho) = \frac{N(N-1)\rho(1-\rho)}{N(N-1)\rho} = 1 - \rho \qquad (4)$$

To quantify the actual asymmetry of an observed network $G_\tau$, we normalize $\phi(G_\tau)$ by the asymmetry that is expected simply due to chance given the density of this network. So, we define the *normalized asymmetry* of $G_\tau$ as

$$\Phi(G_\tau) = \frac{\phi(G_\tau)}{\bar{\phi}(G_\tau)} \qquad (5)$$

which is well defined as long as $\rho(G_\tau) < 1$.

MANIA is based on the following premise: *the inferred directed network should be as symmetric as possible.* The reason is that the tractography process is unable to infer the actual direction (polarity) of the underlying neural fibers. So, if there are some fibers connecting two voxels X and Y, tractography should be able, in principle, to follow this connection in both directions, from X to Y and from Y to X. We do not claim that two connected ROIs are always attached with both afferent and efferent fibers; instead, we argue that tractography is not able to discover the polarity of those fibers and so the corresponding connection should be trace-able in both directions.

The presence of some streamlines from some voxels in ROI X to ROI Y does not necessarily mean however that the network inference method will detect an edge both from X to Y and from Y to X; this also depends on the parameter τ. Given that we aim to minimize the asymmetry of the inferred network, MANIA aims to select the value of $\tau$ that leads to the lowest possible asymmetry.

The corresponding optimization problem can be stated as follows: determine the adjacency matrix $\hat{G} = G_{\tau*}$, where $\tau^*$ is a value of the connectivity threshold that minimizes the normalized asymmetry of $G_\tau$ across all possible values of $\tau$,

$$\tau^* = \arg\ min_{0 < \tau < 1} \Phi(G_\tau) \tag{6}$$

So, MANIA is based on the premise that there is an ideal value (or range of values) of the connectivity threshold that can correctly classify every directed pair of ROIs as either "connection exists" or "connection does not exist". When such a threshold exists, it will result in a completely symmetric network (because a perfectly accurate tractography-based network cannot be asymmetric). On the other hand, if such an ideal threshold does not exist (for instance, it may be that two connected ROIs are too far from each other and tractography cannot "see" their connection, or that it is impossible for streamlines to cross the white matter/grey matter boundary of a certain ROI in one direction but not in the opposite), then MANIA aims to at least minimize the normalized asymmetry metric, even if the resulting network will not be completely symmetric.

If there is more than one value of $\tau$ that results in the same minimum of the normalized asymmetry (potentially zero), MANIA reports the network with the largest density. The rationale behind this tie-breaker is to avoid trivial solutions that include only a subset of the actual network edges. The previous optimization problem can be solved numerically by scanning the range of $\tau$ values with a given resolution.[1] The density of the resulting network is denoted by

$$\rho^* = \rho(G_{\tau*}) \tag{7}$$

As an illustration of the previous method, Figure 2 shows how the network asymmetry (both $\phi$ and $\Phi$) varies with the density $\rho$ as well as the relation between $\rho$ and $\tau$ for the dataset that corresponds to one of the subjects in our case-study.

## 2.5 Threshold-based network inference with post-symmetrization

A common network inference method is to rely on a given connectivity threshold $\tau$, as shown in Equation (1). This threshold is sometimes chosen to achieve a certain network density or to ensure that the network is connected (Y. Li et al., 2009).

Given that tractography cannot detect the direction of inferred edges, the resulting network can be then "post-symmetrized" as follows. Consider two network nodes $i$ and $j$ and suppose that the fraction of streamlines from $i$ to $j$ is denoted by $T_{i,j}$. Suppose that $T_{i,j} > \tau$ but $T_{j,i} < \tau$. We can resolve the conflicting evidence between the two directions of this edge by comparing the ratio $(T_{i,j} - \tau) / (1 - \tau)$ that reflects our confidence that the edge from $i$ to $j$ exists, with the ratio $(\tau - T_{j,i}) / \tau$ that reflects our confidence that the edge from $j$ to $i$ does *not* exist.

This post-symmetrization step is different than MANIA in several ways. First, post-symmetrization relies on an arbitrary connectivity threshold to make all edges symmetric, while MANIA selects the threshold value that minimizes the asymmetry metric $\Phi$. Second, post-symmetrization considers each pair of nodes individually, while MANIA considers the entire network, normalizing the observed asymmetry by the expected asymmetry of a random network of equal density. Third, post-symmetrization always results in a symmetric network, while MANIA may not do so if there is no value of the connectivity threshold value that would result in perfect symmetry. Optionally, post-symmetrization can also be applied on the output of MANIA, if the resulting MANIA network is not completely symmetric.

---

[1] Since we set the number of streamlines to 5000, the minimum resolution is $\tau = \frac{1}{5000} = 0.0002$.

### 2.6 Performance metrics

MANIA can be viewed as a binary classifier: each possible directed edge is classified as *present* or *absent*. We evaluate MANIA based on the following standard metrics for binary classification: the *false positive rate* (or false alarm) $p_f$, and the *false negative rate* (or miss detection) $p_m$. The former is defined as the fraction of absent edges that are incorrectly classified as present, while the latter is defined as the fraction of present edges that are incorrectly classified as absent.

Also, the *Jaccard similarity* between the sets of edges $E(G)$ and $E(\hat{G})$ of the actual network $G$ and the MANIA network $\hat{G}$, respectively, is defined as

$$J(G, \hat{G}) = \frac{|E(G) \cap E(\hat{G})|}{|E(G) \cup E(\hat{G})|} \tag{8}$$

and it varies between zero (no common edges) and one (identical networks).

### 2.7 Optimal threshold-based network inference

We can also compare the network that results from MANIA with the network that would result if we knew the optimal value $\tau^{opt}$ of the connectivity threshold, i.e., the value of τ that maximizes the Jaccard similarity between the inferred network $G_\tau$ and the ground truth network $G$:

$$\tau^{opt} = \arg\ max_{0 < \tau < 1}\ J(G_\tau, G) \tag{9}$$

Even though it is not possible to know this optimal threshold value when analyzing real tractography data, we can easily compute its value (or range of values) in experiments with synthetically generated networks, where the ground-trith network $G$ is known.

## 2.8 Edge ranking and confidence metric in MANIA

The output of MANIA is an unweighted directed network. We can quantify the level of confidence we have in each edge with the following edge ranking scheme.

Let $\rho_\alpha$ be the minimum network density at which edge α is present. If the edge α is from a source X to a target Y, the lower $\rho_\alpha$ is, the higher the fraction of streamlines from X to Y. Consequently, we can rank edges so that we are more confident in the presence of edge α than of edge β if $\rho_\alpha < \rho_\beta$ (see Figure 3).

We define a *confidence metric* for an edge α that is present in the MANIA network (i.e., $\rho_\alpha < \rho^*$) as follows

$$C(\alpha) = \frac{\rho^* - \rho_\alpha}{\rho^*} \tag{10}$$

$C(\alpha)$ varies from 0 (the edge is only marginally present) to 1 (highest confidence that the edge is present). Similarly, if edge α is absent from the MANIA network (i.e., $\rho_\alpha > \rho^*$), its confidence metric is defined as

$$C(\alpha) = \frac{\rho^* - \rho_\alpha}{1 - \rho^*} \tag{11}$$

and $C(\alpha)$ varies from 0 (the edge is only marginally absent) to -1 (highest confidence that the edge is absent).

We also define a confidence metric for a pair of nodes $(X,Y)$, as the arithmetic mean of the confidence metric of the two directed edges between $X$ and $Y$,

$$C(X,Y) = \frac{C(X \to Y) + C(Y \to X)}{2} \tag{12}$$

Note that one of the two edges may be present while the other may be absent. In that case, the confidence of the corresponding node-pair will be less than the confidence of the present edge.

Note that this edge confidence metric is not related to connection "strength" or "quality", and the resulting network is still meant to be interpreted as an unweighted graph.

## 2.9 Group analysis using MANIA

If the objective is to create a single "average network" based on data from several subjects, the question is how to best aggregate the tractography data from the given group. One approach is to average the diffusion MRI data, after transforming them in a standard space. Another approach is to average the fraction of streamlines from a given seed voxel to a given target ROI, across all subjects. These approaches are sensitive to outliers, variations in the diffusion MRI process across subjects, tractography errors and mapping/warping into a standard space. A third approach could be to construct an individual network for each subject, perhaps using MANIA, and then construct an aggregate network only keeping those edges that appear in a large fraction of subjects. This approach requires a group-level threshold to represent the minimum fraction of subjects that should have a connection. For instance, de Reus et al. have proposed a statistically rigorous method to compute such a threshold (de Reus et al., 2013a).

Here, we propose a different group analysis method, referred to as *group-MANIA*, that is based on the aggregation of edge-rankings across subjects. The rank-based nature of this method makes it robust to outliers.

As in the previous section, the edges of a subject can be ranked based on the minimum network density at which an edge first appears. We are more confident in the presence of edge $\alpha$ than of edge $\beta$ if $\rho_\alpha < \rho_\beta$ (see Figure 3).

Suppose that we compute an edge rank vector $R_m$ for each subject $m$, so that the the lowest rank $R_m(1)$ corresponds to the edge for which we are most confident. The number of possible (not necessarily present) directed edges is the same for all subjects: $N(N-1)$ where $N$ is the number of network nodes.

Given a group of size $M$, we have $M$ distinct rank vectors $R_m, \ m = 1 \cdots M$. The computational problem of *rank aggregation* (Schalekamp and van Zuylen, 2009) is to compute an optimal permutation $\hat{R}$ of the $N(N-1)$ possible edges that captures as well as possible the ordering relations in the $M$ input rank vectors. Specifically, the *Kemeny distance* between two rank vectors $R_1$ and $R_2$ is defined as

$$\sum_i \sum_j \delta_{i,j} \left( R_1, R_2 \right) \qquad (13)$$

where $\delta_{i,j}(R_1, R_2) = 1$ if $R_1$ and $R_2$ disagree in the relative position of elements $i$ and $j$, and zero otherwise. Rank aggregation aims to compute a vector $\hat{R}$ that minimizes the cumulative Kemeny distance between $\hat{R}$ and all input rank vectors $R_m$. It is an NP-hard problem, and so it is typically solved heuristically. We use the Quicksort algorithm (Ailon et al., 2008) since it is has been shown to provide a good approximation of the optimum solution. QuickSort selects a random edge as pivot at each recursive step, while the remaining edges are separated in a left and right list. The left list includes edges that have a lower rank than the pivot in the majority of the subjects; similarly for the right list. The algorithm proceeds recursively in the left and right lists until all edges are ordered.

After computing the optimal aggregate rank vector, we apply MANIA on $\hat{R}$ to compute the network with the minimum normalized asymmetry (as in the case of a single subject). Note however that the input to MANIA in this case is an ordered list of edges $\hat{R}$ rather than the set of connectivity matrices $T$ (see section 2.2). Group-MANIA forms a network with the *first* $K = N(N-1)\rho$ edges in $\hat{R}$, and it computes the normalized asymmetry of that network. It then repeats this step, for all values of K, to identify the network with the minimum value of $\Phi$. We refer to the resulting network as the *rank-aggregated network*.

## 2.10 Synthetically generated networks

To evaluate the accuracy and sensitivity of MANIA in a reliable manner we need to rely on synthetic networks rather than actual DTI and tractography data. The benefit of these computational experiments is that we can test MANIA under a wide range of noise conditions and for arbitrary network densities. Unfortunately there are no good statistical models for the noise in DTI and tractography data (Jbabdi et al., 2011). We evaluate MANIA based on a simple noise model that is based on the theory of *maximum entropy distributions*, as described next.

For simplicity, each ROI of the synthetically generated networks is simply a voxel. Modeling multi-voxel ROIs in these simulation experiments would not add any new insights. Suppose that the directed network between $N$ nodes is represented by the $N \times N$ adjacency matrix $G$. Let $T_{i,j}$ be the fraction of streamlines that originate from node $i$ and terminate at node $j$. Ideally, in the absence of any noise in the DTI data and without any errors in the tractography process, it should be that

$$T_{i,j} = \begin{cases} 1 & if\ G_{i,j} = 1\ or\ G_{j,i} = 1 \\ 0 & if\ G_{i,j} = 0\ and\ G_{j,i} = 0 \end{cases} \tag{14}$$

So, if there is an edge between nodes $i$ and $j$ in either direction, the fraction of streamlines from node $i$ to node $j$ should be 100%; otherwise, it should be zero.

In practice, there is significant noise in DTI data and the tractography process can be error-prone, especially when the ROIs are in grey matter and/or when neural fibers cross each other, split or merge, or fan out as they approach their targets. Consequently, the tractography output may show that some streamlines do not reach from node $i$ to node $j$ even when the two nodes are connected, or that some streamlines get from $i$ to $j$ even when there is no connection between the two nodes. We model these errors probabilistically, as follows:

$$T_{i,j} = \begin{cases} 1 - Z_1 & if\ G_{i,j} = 1\ or\ G_{j,i} = 1 \\ Z_2 & if\ G_{i,j} = 0\ and\ G_{j,i} = 0 \end{cases} \tag{15}$$

where $Z_1$ and $Z_2$ are two (generally different) random variables with [0,1] support. If their probability mass is concentrated close to 0, the results of the tractography process are not significantly affected by noise. On the other hand, if these two random variables are uniformly distributed in [0,1], the tractography results are completely random and any network inference process is hopeless.

We model the two random variables $Z_1$ and $Z_2$ with the *Maximum Entropy distribution with one-degree of freedom.* In this case, this distribution is the truncated exponential distribution with support [0,1],

$$f_Z(\zeta) = \begin{cases} \dfrac{\alpha}{1 - e^{-\alpha}} e^{-\alpha\zeta} & \zeta \in [0,1] \\ 0 & otherwise, \end{cases} \quad (16)$$

where $\alpha > 0$ is a parameter that determines the mean and variance of the distribution. Instead of controlling $\alpha$, we control the intensity of noise through the mean of $Z$,

$$\mu = E[Z] = \frac{1 - (1 + \alpha)e^{-\alpha}}{\alpha(1 - e^{-\alpha})} \quad (17)$$

The two distributions $Z_1$ and $Z_2$ follow this statistical model with means $\mu_1$ and $\mu_2$ respectively. Figure 4 shows the previous distribution for four values of $\mu$. Note that the distribution $Z$ becomes almost "flat" (close to the uniform distribution) when its mean is higher than 0.3, meaning that tractography would be extremely inaccurate when the noise intensity exceeds that level. In the rest of this paper we limit the range of $\mu_1$ and $\mu_2$ between 0 and 0.3.

## 3  Results

### 3.1  Evaluation with synthetic data

We evaluate MANIA based on computational experiments with synthetic data and random networks. The "ground-truth" networks $G$ are constructed as follows. Suppose that $G$ has $N$ nodes and density $\rho$. We place $\left\lfloor \rho \frac{N(N-1)}{2} \right\rfloor$ undirected edges between randomly selected but distinct pairs of nodes. Note that $G$ is symmetric by construction because the tractography process cannot infer the true polarity of the underlying neural fibers. Given $G$, we then create the tractography matrix $T$ that represents the "noisy" fraction of streamlines between any pair of nodes, as shown in equation (15). Note that the fraction of streamlines from a node X to a node Y is typically different than the fraction of streamlines from Y to X. In the following experiments, $N$ is set to 50 nodes, and each experiment is repeated for 1000 networks $G$.

We first examine the effect of post-symmetrization on the accuracy of both threshold-based network inference and MANIA. In the former, the connections are determined based on a given threshold $\tau_0$, as discussed in Section 2.5. We denote the Jaccard similarity between the inferred network and the ground-truth network with $J_{SYM}$ when post-symmetrization is performed, and with $J_{NO-SYM}$ otherwise. Figure 5 shows the difference $\Delta J = J_{SYM} - J_{NO-SYM}$ for several choices of $\tau_0$ as well as for MANIA. Each box-plot is generated from 1000 experiments; in each experiment we generate a random network with density between 0 and 1, while the noise parameters $\mu_1$ and $\mu_2$ are uniformly distributed between 0 and 0.3. The red line corresponds to the median, the box boundaries correspond to the 25th and 75th percentiles, while the dashed lines show the 10th and 90th percentiles. In all cases, $\Delta J > 0$ (one-sided Mann-Whitney U test – p-values shown next to each box plot), meaning that *post-symmetrization helps to improve the accuracy of network inference*. This is true for both MANIA and threshold-based inference, even though the improvement is larger for the latter. Because of the positive effect of post-symmetrization, in the rest of the paper we apply it in all network inference experiments.

Figure 6 illustrates the performance of MANIA in the two-dimensional space defined by the noise parameters $\mu_1$ and $\mu_2$ for three values of the network density. Each square in these heat maps is the median across 1000 experiments. The false positive and false negative rates are close to 0 (less than 5%) in the lower-left half of each heat map (i.e., when $\mu_1 + \mu_2 < 0.3$). For higher values of the noise intensity, the accuracy of MANIA depends on the density of the underlying network. In the case of sparse networks, MANIA also infers a sparse network and most errors are false negatives, i.e., MANIA does not detect some of the few existing edges. For dense networks, MANIA also infers a dense network and most errors are false positives, i.e., MANIA detects a few extra edges that do not actually exist. In mid-range densities, the errors are more balanced between false positives and false negatives. In all cases the maximum false positive (or negative) rate when $\mu_1 = \mu_2 = 0.3$ is less than 25%. Recall from Figure 4 that these noise intensity levels should be considered quite high in practice.

Figure 7 compares the MANIA-inferred network with the network that corresponds to the optimal threshold $\tau^{opt}$, as discussed in Section 2.7. Specifically, the heat maps of Figure 7 compare the Jaccard similarity $J_{MANIA}$ between MANIA and the ground-truth network, with the Jaccard similarity $J_{\tau^{opt}}$ between the optimal threshold based network and the ground-truth network. The accuracy of MANIA is typically close to that of the optimal threshold method. Even under the highest noise intensity we consider ($\mu_1 = \mu_2 = 0.3$), $J_{MANIA}$ is only 10% lower than $J_{\tau^{opt}}$. These results suggest that MANIA selects automatically a connectivity threshold value that results in almost optimal accuracy, across all possible such threshold values.

Finally, Figure 8 compares MANIA with five given threshold values $\tau_0$. The comparison is in terms of the Jaccard similarity difference $\Delta J = J_{MANIA} - J_{\tau_0}$. As in Figure 5, each box-plot is generated from 1000 experiments in which we vary the network density between 0 and 1, and the noise parameters $\mu_1$ and $\mu_2$ between 0 and 0.3. The median $\Delta J$ is always positive and the distribution of $\Delta J$ is skewed towards positive values (one-sided Mann-Whitney U test – p-values shown next to each box plot), meaning that MANIA typically performs better than a fixed threshold scheme, independent of the selected threshold.

## 3.2  Case-study: a rank-aggregated network between 18 ROIs

We applied MANIA in the DTI data presented in Section 2.1, based on the 18 ROIs listed in Table 1. A single rank-aggregated network is constructed, using the group analysis method of Section 2.9, aggregating data from 28 subjects. The rank-aggregated network is shown in Figure 9.

Two ROIs (Pc and BA40) appear to not be *directly* connected with the other 16 ROIs; of course there may be indirect connections through other ROIs that have not been included here (we return to this point in Section 4). Every edge in the connected component of Figure 9 has been detected in both directions, i.e., MANIA identifies a completely symmetric network in this case (i.e., no post-symmetrization is needed). The density of the connected component (16 ROIs) is 19%. The color of each edge in Figure 9 represents the fraction of the 28 subjects that have the corresponding edge in their individual networks (constructed by MANIA).

We measured the "centrality" of each node in the rank-aggregated network, based on four centrality metrics (degree, closeness, betweenness, PageRank) (Newman, 2010). Different centrality metrics focus on different notions of importance. For instance, the degree centrality metric associates importance with the number of direct connections a node has; BA32 (Ventral anterior cingulate) has the largest number (six) of direct connections in this network (see Table 2). This may be because BA32 is spatially adjacent to both BA10 and BA25, and those ROIs are also of high degree. The betweenness centrality of a node X, on the other hand, focuses on the number of shortest paths between any pair of nodes that go through X; BA25 (subcallosal cingulate) is the most important node from this perspective because it serves as the "unique bridge" between the 6 red nodes at its left and the 9 blue nodes at its right. BA25 is also the most central node in terms of its average distance to all other nodes (closeness centrality).

Similarly, we measured the edge centrality of all connected node pairs. In terms of edge betweenness centrality, the connection between BA25 and the Nucleus Accumbens (Acb) is by far the most central in this network. It is interesting to note that this edge includes the segment of white matter that is the target of Deep Brain Stimulation (DBS) therapies for the treatment of MDD (Mayberg et al., 2005). In fact, the DBS target is typically the point at which the fibers between (BA25-Acb), (BA25-BA32) and (BA25-BA24) intersect.

Figure 10 shows some percentiles of the per-subject node-pair confidence metric (median, 25-75th percentiles, 10-90th percentiles, and outliers) for each node-pair that appears connected in at least one of the 28 subjects. The connections between the following node pairs appear in all subjects and have the highest confidence: Hp-Acb, Amg-Acb, BA47-Ins. On the other hand, the following connections appear only in some subjects and their confidence metric varies around zero: Th-BS, BA46-BA9, BA6-Ins. Some connections that appear in 1-2 subjects but have very low confidence are: Pc-BA24, BA11-BA24, Ins-BA25, BA40-BA6.

# 4 Discussion

There is an ongoing debate about the accuracy of tractography-based structural network inference. For instance, Thomas et al. have shown that inferring long-range anatomical connections between grey matter ROIs from DWI data is inherently inaccurate (Thomas et al., 2014). The authors also note that *"(probabilistic tractography methods) are less susceptive to changes in the composition of an ROI but only if an optimized threshold can be derived and used."* More recently Reveley et al. (Reveley et al., 2015) have investigated the key reasons behind the negative results of (Thomas et al., 2014). They showed that the dense system of white matter fibers residing just under the cortical sheet poses severe challenges for long-range tractography, concluding that it is *"extremely difficult to determine precisely where small axonal tracts join and leave larger white matter fasciculi."* Another critique, by Jones et al., argued that the number of reconstructed streamlines (i.e., the Number of Streamlines, or NoS, in deterministic tractography) should not be viewed as equivalent to "fiber count", and similarly, the connectivity likelihoods inferred from probabilistic tractography should not be viewed as "connection strengths" (Jones et al., 2013).

In light of the previous results, we believe that there is a need for new network inference methods. MANIA is moving in the right direction for the following reasons:

a) The results of (Thomas et al., 2014) suggest that probabilistic tractography can be more accurate than deterministic tractography in terms of sensitivity and specificity as long as its parameters are appropriately optimized. MANIA is indeed mostly applicable to probabilistic tractography, and its main focus is how to "self-configure" its connectivity threshold $\tau$ in an optimized manner, relying on what we expect to be true about the structure of the resulting solution (namely, a symmetric network).

b) The results of (Reveley et al., 2015) suggest that it is risky to artificially dilate the given ROIs, which are typically mostly grey matter, so that they also include some white matter voxels. Those voxels may be part of the white matter fiber systems that reside just under the cortical sheet. In other words, if our goal is to understand the connectivity between grey matter ROIs, we should not use tractography seeds that reside in white matter; instead, we need to seed from grey matter even if the diffusion signal is much weaker there. So, we need to expect that some connections may appear as asymmetric, which is what MANIA anticipates.

c) MANIA does not construct weighted networks. As previously discussed, it is debatable whether the results of tractography (NoS or fractions of connected streamlines) can be viewed as a proxy for "fiber count" or "connection strength".

d) The results of (Reveley et al., 2015) can be interpreted as follows: because it is hard for any tractography method to accurately cross the white matter/grey matter boundary (WGB), especially in the case of cortical ROIs, a network inference method should be able to deal somehow with erroneous measurements about specific connections. In other words, just because tractography failed to cross the WGB going from seed X to target Y does *not* mean that we should conclude that X and Y are not connected. And so, given that the input data about individual connections is quite noisy, we need to examine if there is any additional "hidden structure" in the inference problem that we can exploit. If this is case, we can then look for a solution that satisfies the constraints of that additional structure. In MANIA, this "hidden structure" in the inference problem is that the resulting network should be as symmetric as possible.

Of course, we do *not* claim that MANIA addresses every concern about tractography-based network inference. On the contrary, there are more open issues that need to be addressed. Two of them are further discussed next.

Even if the thresholding problem is adequately addressed with MANIA, there is another important problem in structural network inference: the *distance bias* of the tractography process (L. Li et al., 2012a). It is harder to discover connections between distal regions due to the accumulation of uncertainty in long streamlines, causing false negatives for long-range connections (L. Li et al., 2012b). Additionally, it is more likely to incorrectly detect connections between proximal regions, especially in the presence of crossing or turning fibers, causing false positives. The FDT toolbox provides a "distance correction" option by multiplying the number of streamlines that cross a voxel by the average length of those streamlines[1] – there is no evidence however that this simple form of distance correction is able to improve significantly the accuracy of the network inference process (L. Li et al., 2012b). A more sophisticated method is that of (Morris et al., 2008), which compares the tractography-generated connectivity probabilities with a null model that gives the corresponding connectivity probabilities with a random tracking process that is dominated by the same distance effects. We view distance correction methods as an independent processing step that can be applied prior to applying MANIA.

---

[1] http://fsl.fmrib.ox.ac.uk/fsl/fslwiki/FDT/UserGuide

For instance, the method by Morris et al. first creates a "null frequency of connection map", it then filters the "experimental frequency of connection map" that is produced by a probabilistic tractography tool, resulting in the so-called "significance of connection map" (which is supposed to have fewer false positives). MANIA can be then applied on the latter, rather than on the experimental frequency of connection map. Even though it is still not clear if the Morris distance correction method is sufficient to completely address the distance correction bias (Taljan et al., 2011), we anticipate that the combination of MANIA with the distance correction method of Morris et al. will improve the accuracy of the resulting networks.

A network representation consists of both nodes and edges. The clinical and research value of representing a brain as a network depends critically on the selected nodes and on the exact boundaries of the corresponding ROIs (Zalesky et al., 2010). MANIA assumes that the set of given nodes is sufficiently specified, and that their spatial boundaries are accurately defined. In practice, this step of the network inference process is always an "inexact science" given that the functional role of any given ROI is at best only partially understood and the anatomical boundary of each ROI is subject-dependent (Tzourio-Mazoyer et al., 2002).

A voxel-level analysis (van den Heuvel et al., 2008) avoids the selection of functionally specific ROIs but it makes it harder to associate the topological properties of the observed network, which now consists of many thousands of nodes, to any known brain circuits and their function. Again, we view this important issue as orthogonal to MANIA: improved brain parcellation methods, such as data-driven parcellations (Power et al., 2011; Yeo et al., 2011) and decreasing voxel sizes can be used jointly with MANIA to identify structural networks that are consistent, or that can explain well, the observed spatio-temporal correlations in resting-state or task-based fMRI analyses. This coupled exploitation of fMRI and diffusion MRI data has provided valuable insights about the underlying anatomy of the brain structures that result in the Default Mode Network (Buckner et al., 2008), and they can become more common now that the HCP project provides both functional and diffusion data for hundreds of subjects (Van Essen et al., 2013).

In our 18-ROI case-study, summarized in Figure 9, the use of mostly large ROIs that do not necessarily correspond to distinct functional units, together with the distance bias of the tractography process, may account for the lack of certain expected connections. Two such expected connections are between Pc and BA40 (Greicius et al., 2009), and between BA9 and BA40 (Petrides and Pandya, 2007); the latter is a long-distance connection. Additionally, large cortical ROIs such as BA9 and BA40 are only imprecisely defined, which may also explain the absence of some of their connections. The limbic and subcortical ROIs, on ther other hand, are more precisely defined and their connections are mostly running over shorter distances. These findings suggest that MANIA should be evaluated in the future jointly with, first, advanced distance correction methods, and second, with either more precisely defined ROIs or on a whole-brain parcellation template.

## Acknowledgments

C.D. and S.B. are grateful to Prof. Olaf Sporns and his group at Indiana University for hosting S.B. at the Sporns lab during the summer of 2011 and for discussions about early stages of this work. Also, the authors would like to thank Dr. Jim Rilling for his contribution of the DTI and structural MRI data that served as basis for the analysis presented in this paper.

## Disclosure Statement

Every co-author of this paper declares that he/she has no competing financial interests.

**References**

Ailon, N., Charikar, M., Newman, A. (2008). Aggregating inconsistent information: ranking and clustering. *Journal of the ACM (JACM), 55*(5), 23.

Bassett, D. S., Brown, J. A., Deshpande, V., Carlson, J. M., Grafton, S. T. (2011). Conserved and variable architecture of human white matter connectivity. *NeuroImage, 54*(2), 1262-1279.

Bassett, D. S., Bullmore, E., Verchinski, B. A., Mattay, V. S., Weinberger, D. R., Meyer-Lindenberg, A. (2008). Hierarchical organization of human cortical networks in health and schizophrenia. *The Journal of Neuroscience, 28*(37), 9239-9248.

Bastiani, M., Shah, N. J., Goebel, R., Roebroeck, A. (2012). Human cortical connectome reconstruction from diffusion weighted MRI: the effect of tractography algorithm. *NeuroImage, 62*(3), 1732-1749.

Behrens, T., Berg, H. J., Jbabdi, S., Rushworth, M., Woolrich, M. (2007). Probabilistic diffusion tractography with multiple fibre orientations: What can we gain? *NeuroImage, 34*(1), 144-155.

Behrens, T., Woolrich, M., Jenkinson, M., Johansen-Berg, H., Nunes, R., Clare, S., et al. (2003). Characterization and propagation of uncertainty in diffusion-weighted MR imaging. *Magnetic Resonance in Medicine, 50*(5), 1077-1088.

Beucke, J. C., Sepulcre, J., Eldaief, M. C., Sebold, M., Kathmann, N., Kaufmann, C. (2014). Default mode network subsystem alterations in obsessive-compulsive disorder. *The British Journal of Psychiatry, 205*(5), 376-382.

Blumensath, T., Jbabdi, S., Glasser, M. F., Van Essen, D. C., Ugurbil, K., Behrens, T. E., et al. (2013). Spatially constrained hierarchical parcellation of the brain with resting-state fMRI. *NeuroImage, 76*, 313-324.

Buckner, R. L., Andrews-Hanna, J. R., Schacter, D. L. (2008). The brain's default network. *Annals of the New York Academy of Sciences, 1124*(1), 1-38.

Buckner, R. L., Snyder, A. Z., Shannon, B. J., LaRossa, G., Sachs, R., Fotenos, A. F., et al. (2005). Molecular, structural, and functional characterization of Alzheimer's disease: evidence for a relationship between default activity, amyloid, and memory. *The Journal of Neuroscience, 25*(34), 7709-7717.

Cheng, H., Wang, Y., Sheng, J., Sporns, O., Kronenberger, W. G., Mathews, V. P., et al. (2012). Optimization of seed density in DTI tractography for structural networks. *Journal of neuroscience methods, 203*(1), 264-272.

Chen, X., Errangi, B., Li, L., Glasser, M. F., Westlye, L. T., Fjell, A. M., et al. (2013). Brain aging in humans, chimpanzees (Pan troglodytes), and rhesus macaques (Macaca mulatta): magnetic resonance imaging studies of macro-and microstructural changes. *Neurobiology of aging, 34*(10), 2248-2260.

Ciccarelli, O., Catani, M., Johansen-Berg, H., Clark, C., Thompson, A. (2008). Diffusion-based tractography in neurological disorders: concepts, applications, and future developments. *The Lancet Neurology, 7*(8), 715-727.

Craddock, R. C., Holtzheimer, P. E., Hu, X. P., Mayberg, H. S. (2009). Disease state prediction from resting state functional connectivity. *Magnetic Resonance in Medicine, 62*(6), 1619-1628.

Craddock, R. C., James, G. A., Holtzheimer, P. E., Hu, X. P., Mayberg, H. S. (2012). A whole brain fMRI atlas generated via spatially constrained spectral clustering. *Human brain mapping, 33*(8), 1914-1928.

Daianu, M., Jahanshad, N., Nir, T. M., Toga, A. W., Jack Jr, C. R., Weiner, M. W., et al. (2013). Breakdown of brain connectivity between normal aging and Alzheimer's disease: a structural k-core network analysis. *Brain connectivity, 3*(4), 407-422.

Damoiseaux, J. S., Greicius, M. D. (2009). Greater than the sum of its parts: a review of studies combining structural connectivity and resting-state functional connectivity. *Brain Structure and Function, 213*(6), 525-533.

de Reus, M. A., van den Heuvel, M. P. (2013a). Estimating false positives and negatives in brain networks. *NeuroImage, 70*, 402-409.

de Reus, M. A., Van den Heuvel, M. P. (2013b). The parcellation-based connectome: limitations and extensions. *NeuroImage, 80*, 397-404.

Descoteaux, M., Deriche, R., Knösche, T. R., Anwander, A. (2009). Deterministic and probabilistic tractography based on complex fibre orientation distributions. *Medical Imaging, IEEE Transactions on, 28*(2), 269-286.

Duda, J. T., Cook, P. A., Gee, J. C. (2014). Reproducibility of graph metrics of human brain structural networks. *Frontiers in neuroinformatics, 8*.

Elman, I., Borsook, D., Volkow, N. D. (2013). Pain and suicidality: insights from reward and addiction neuroscience. *Progress in neurobiology, 109*, 1-27.

Fornito, A., Bullmore, E. T. (2015). Connectomics: a new paradigm for understanding brain disease. *European Neuropsychopharmacology, 25*(5), 733-748.

Gong, G., He, Y., Concha, L., Lebel, C., Gross, D. W., Evans, A. C., et al. (2009). Mapping anatomical connectivity patterns of human cerebral cortex using in vivo diffusion tensor imaging tractography. *Cerebral Cortex, 19*(3), 524-536.

Greicius, M. D., Supekar, K., Menon, V., Dougherty, R. F. (2009). Resting-state functional connectivity reflects structural connectivity in the default mode network. *Cerebral Cortex, 19*(1), 72-78.

Hagmann, P., Cammoun, L., Gigandet, X., Meuli, R., Honey, C. J., Wedeen, V. J., et al. (2008). Mapping the structural core of human cerebral cortex. *PLoS Biol, 6*(7), e159.

Hagmann, P., Kurant, M., Gigandet, X., Thiran, P., Wedeen, V. J., Meuli, R., et al. (2007). Mapping human whole-brain structural networks with diffusion MRI. *PloS one, 2*(7), e597.

James, G. A., Kelley, M. E., Craddock, R. C., Holtzheimer, P. E., Dunlop, B. W., Nemeroff, C. B., et al. (2009). Exploratory structural equation modeling of resting-state fMRI: applicability of group models to individual subjects. *NeuroImage, 45*(3), 778-787.

Jbabdi, S., Johansen-Berg, H. (2011). Tractography: where do we go from here? *Brain connectivity, 1*(3), 169-183.

Jenkinson, M., Smith, S. (2001). A global optimisation method for robust affine registration of brain images. *Medical image analysis, 5*(2), 143-156.

Jones, D. K., Knösche, T. R., Turner, R. (2013). White matter integrity, fiber count, and other fallacies: the do's and don'ts of diffusion MRI. *NeuroImage, 73*, 239-254.

Lancaster, J. L., Woldorff, M. G., Parsons, L. M., Liotti, M., Freitas, C. S., Rainey, L., et al. (2000). Automated Talairach atlas labels for functional brain mapping. *Human brain mapping, 10*(3), 120-131.

Li, L., Rilling, J. K., Preuss, T. M., Glasser, M. F., Damen, F. W., Hu, X. (2012a). Quantitative assessment of a framework for creating anatomical brain networks via global tractography. *NeuroImage, 61*(4), 1017-1030.

Li, L., Rilling, J. K., Preuss, T. M., Glasser, M. F., Hu, X. (2012b). The effects of connection reconstruction method on the interregional connectivity of brain networks via diffusion tractography. *Human brain mapping, 33*(8), 1894-1913.

Li, Y., Liu, Y., Li, J., Qin, W., Li, K., Yu, C., et al. (2009). Brain anatomical network and intelligence. *PLoS computational biology, 5*(5), e1000395.

Mayberg, H. S. (1997). Limbic-cortical dysregulation: a proposed model of depression. *The Journal of neuropsychiatry and clinical neurosciences*.

Mayberg, H. S., Lozano, A. M., Voon, V., McNeely, H. E., Seminowicz, D., Hamani, C., et al. (2005). Deep brain stimulation for treatment-resistant depression. *Neuron, 45*(5), 651-660.

McIntosh, A. M., Maniega, S. M., Lymer, G. K. S., McKirdy, J., Hall, J., Sussmann, J. E., et al. (2008). White matter tractography in bipolar disorder and schizophrenia. *Biological psychiatry, 64*(12), 1088-1092.

McKeown, M. J., Makeig, S., Brown, G. G., Jung, T.-P., Kindermann, S. S., Bell, A. J., et al. (1997). Analysis of fMRI data by blind separation into independent spatial components: DTIC Document.

Mori, S., Crain, B. J., Chacko, V., Van Zijl, P. (1999). Three-dimensional tracking of axonal projections in the brain by magnetic resonance imaging. *Annals of neurology, 45*(2), 265-269.

Mori, S., van Zijl, P. (2002). Fiber tracking: principles and strategies-a technical review. *NMR in Biomedicine, 15*(7-8), 468-480.

Morris, D. M., Embleton, K. V., Parker, G. J. (2008). Probabilistic fibre tracking: differentiation of connections from chance events. *NeuroImage, 42*(4), 1329-1339.

Newman, M. (2010). *Networks: an introduction*: Oxford University Press.

Parker, G. J., Haroon, H. A., Wheeler-Kingshott, C. A. (2003). A framework for a streamline-based probabilistic index of connectivity (PICo) using a structural interpretation of MRI diffusion measurements. *Journal of magnetic resonance imaging, 18*(2), 242-254.

Peterson, A., Thome, J., Frewen, P., Lanius, R. A. (2014). Resting-state neuroimaging studies: a new way of identifying differences and similarities among the anxiety disorders? *Canadian journal of psychiatry. Revue canadienne de psychiatrie, 59*(6), 294.

Petrides, M. (2005). Lateral prefrontal cortex: architectonic and functional organization. *Philosophical Transactions of the Royal Society B: Biological Sciences, 360*(1456), 781-795.

Petrides, M., Pandya, D. N. (2007). Efferent association pathways from the rostral prefrontal cortex in the macaque monkey. *The Journal of Neuroscience, 27*(43), 11573-11586.

Power, J. D., Cohen, A. L., Nelson, S. M., Wig, G. S., Barnes, K. A., Church, J. A., et al. (2011). Functional network organization of the human brain. *Neuron, 72*(4), 665-678.

Reese, T., Heid, O., Weisskoff, R., Wedeen, V. (2003). Reduction of eddy-current-induced distortion in diffusion MRI using a twice-refocused spin echo. *Magnetic Resonance in Medicine, 49*(1), 177-182.

Reveley, C., Seth, A. K., Pierpaoli, C., Silva, A. C., Yu, D., Saunders, R. C., et al. (2015). Superficial white matter fiber systems impede detection of long-range cortical connections in diffusion MR tractography. *Proceedings of the National Academy of Sciences, 112*(21), E2820-E2828.

Robinson, E. C., Hammers, A., Ericsson, A., Edwards, A. D., Rueckert, D. (2010). Identifying population differences in whole-brain structural networks: a machine learning approach. *NeuroImage, 50*(3), 910-919.

Rubinov, M., Sporns, O. (2010). Complex network measures of brain connectivity: uses and interpretations. *NeuroImage, 52*(3), 1059-1069.

Schalekamp, F., van Zuylen, A. (2009). Rank Aggregation: Together We're Strong. *ALENEX* (pp. 38-51).

Seminowicz, D., Mayberg, H., McIntosh, A., Goldapple, K., Kennedy, S., Segal, Z., et al. (2004). Limbic-frontal circuitry in major depression: a path modeling metanalysis. *NeuroImage, 22*(1), 409-418.

Sporns, O. (2012). *Discovering the human connectome*: MIT press.

Sporns, O. (2013). Making sense of brain network data. *Nature methods, 10*(6), 491-493.

Sporns, O., Tononi, G., Kötter, R. (2005). The human connectome: a structural description of the human brain. *PLoS computational biology, 1*(4), e42.

Taljan, K., McIntyre, C., Sakaie, K. (2011). Anatomical Connectivity Between Subcortical Structures. *Brain connectivity, 1*(2), 111-118.

Thirion, B., Varoquaux, G., Dohmatob, E., Poline, J.-B. (2014). Which fMRI clustering gives good brain parcellations? *Frontiers in neuroscience, 8*.

Thomas, C., Frank, Q. Y., Irfanoglu, M. O., Modi, P., Saleem, K. S., Leopold, D. A., et al. (2014). Anatomical accuracy of brain connections derived from diffusion MRI tractography is inherently limited. *Proceedings of the National Academy of Sciences, 111*(46), 16574-16579.

Thompson, P. M., Schwartz, C., Lin, R. T., Khan, A. A., Toga, A. W. (1996). Three-dimensional statistical analysis of sulcal variability in the human brain. *The Journal of Neuroscience, 16*(13), 4261-4274.

Tzourio, N., Petit, L., Mellet, E., Orssaud, C., Crivello, F., Benali, K., et al. (1997). Use of anatomical parcellation to catalog and study structure-function relationships in the human brain. *Human brain mapping, 5*(4), 228-232.

Tzourio-Mazoyer, N., Landeau, B., Papathanassiou, D., Crivello, F., Etard, O., Delcroix, N., et al. (2002). Automated anatomical labeling of activations in SPM using a macroscopic anatomical parcellation of the MNI MRI single-subject brain. *NeuroImage, 15*(1), 273-289.

van den Heuvel, M. P., Stam, C. J., Boersma, M., Pol, H. H. (2008). Small-world and scale-free organization of voxel-based resting-state functional connectivity in the human brain. *NeuroImage, 43*(3), 528-539.

Van Essen, D. C., Smith, S. M., Barch, D. M., Behrens, T. E., Yacoub, E., Ugurbil, K., et al. (2013). The WU-Minn human connectome project: an overview. *NeuroImage, 80*, 62-79.

Van Wijk, B. C., Stam, C. J., Daffertshofer, A. (2010). Comparing brain networks of different size and connectivity density using graph theory. *PloS one, 5*(10), e13701.

Yeo, B. T., Krienen, F. M., Sepulcre, J., Sabuncu, M. R., Lashkari, D., Hollinshead, M., et al. (2011). The organization of the human cerebral cortex estimated by intrinsic functional connectivity. *Journal of neurophysiology, 106*(3), 1125-1165.

Zalesky, A., Fornito, A., Harding, I. H., Cocchi, L., Yücel, M., Pantelis, C., et al. (2010). Whole-brain anatomical networks: does the choice of nodes matter? *NeuroImage, 50*(3), 970-983.

**Fig1.** Running tractography with streamlines from a seed voxel $q$ in the $i$'th ROI to the $j$'th target ROI. Two streamlines hit the target ROI, therefore $T_i(q, j) = \frac{2}{3}$.

**Fig2.** Top: Network density $\rho$ as a function of the connectivity threshold $\tau$ (plotted for one subject in our DTI dataset). Bottom: Network asymmetry (red) and normalized network asymmetry (blue) as functions of the network density $\rho$. The optimal density $\rho^*$ is the largest value that minimizes the normalized network asymmetry.

**Fig3.** As we decrease the connectivity threshold, each edge first appears at a certain value of the network density. If this density is larger than $\rho^*$, the corresponding edge is *not* present in the MANIA network.

**Fig4.** Probabilistic error model (Z) for tractography-generated connection probabilities using the maximum entropy distribution (with one degree of freedom).

**Fig5.** The Jaccard similarity difference ($\Delta J = J_{SYM} - J_{NO-SYM}$) with and without post-symmetrization for five threshold values and for MANIA. Each box plot is generated from 1000 experiments; in each experiment we generate a random network with density between 0 and 1, while the noise parameters $\mu_1$ and $\mu_2$ are uniformly distributed between 0 and 0.3. The red line corresponds to the median, the box boundaries to the 25th and 75th percentiles, while the dashed lines show the 10th and 90th percentiles. In all cases, $\Delta J > 0$ (one-sided Mann-Whitney U test – p-values are shown next to each box plot) meaning that *post-symmetrization helps to improve the accuracy of network inference*.

**Fig6.** False positive rate and false negative rate of MANIA as a function of $\mu_1$ and $\mu_2$ for sparse networks ($\rho_G = 0.1$), medium density networks ($\rho_G = 0.5$) and dense networks ($\rho_G = 0.9$). Each square is the average of 1000 independent simulations.

**Fig7.** The Jaccard similarity difference ($\Delta J = J_{\tau^{opt}} - J_{MANIA}$) between MANIA and the optimal threshold-based scheme as a function of $\mu_1$ and $\mu_2$ for sparse networks ($\rho_G = 0.1$), medium density networks ($\rho_G = 0.5$) and dense networks ($\rho_G = 0.9$). Each square is the average of 1000 independent simulations.

**Fig8.** The Jaccard similarity difference ($\Delta J = J_{MANIA} - J_{\tau_0}$) between MANIA and five given threshold values. The accuracy comparisons are made after post-symmetrization. Each box plot is generated from 1000 experiments; in each experiment we generate a random network with density between 0 and 1, while the noise parameters $\mu_1$ and $\mu_2$ are uniformly distributed between 0 and 0.3. The red line corresponds to the median, the box boundaries to the 25th and 75th percentiles, while the dashed lines show the 10th and 90th percentiles. In all cases, $\Delta J > 0$ (one-sided Mann-Whitney U test – p-values are shown next to each box plot) meaning that *MANIA is more accurate than inferring the network based on a fixed threshold*.

**Fig9.** The rank-aggregated network, based on DTI data from 28 subjects, between the 18 ROIs in Table 1. Every edge in the connected component has been detected in both directions, i.e., MANIA identifies a completely symmetric network (no post-symmetrization is needed). The density of the connected component is 19%. The color of each edge represents the fraction of subjects that have the corresponding edge in their individual MANIA-based networks.

**Fig10.** Several percentiles of the node-pair confidence metric for each node-pair that appears connected in at least one of the 28 subjects.

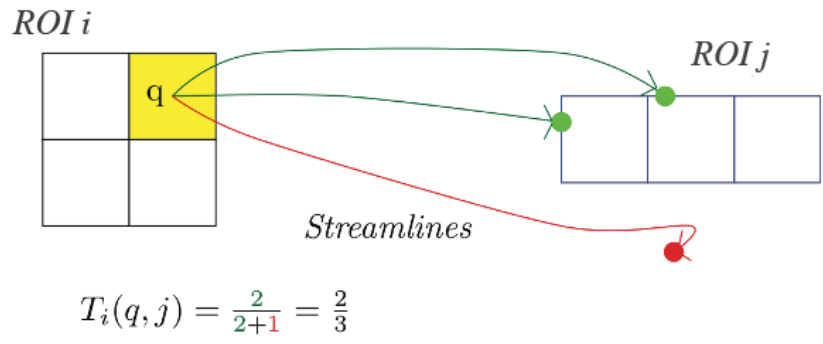

$$T_i(q, j) = \frac{2}{2+1} = \frac{2}{3}$$

Fig.1

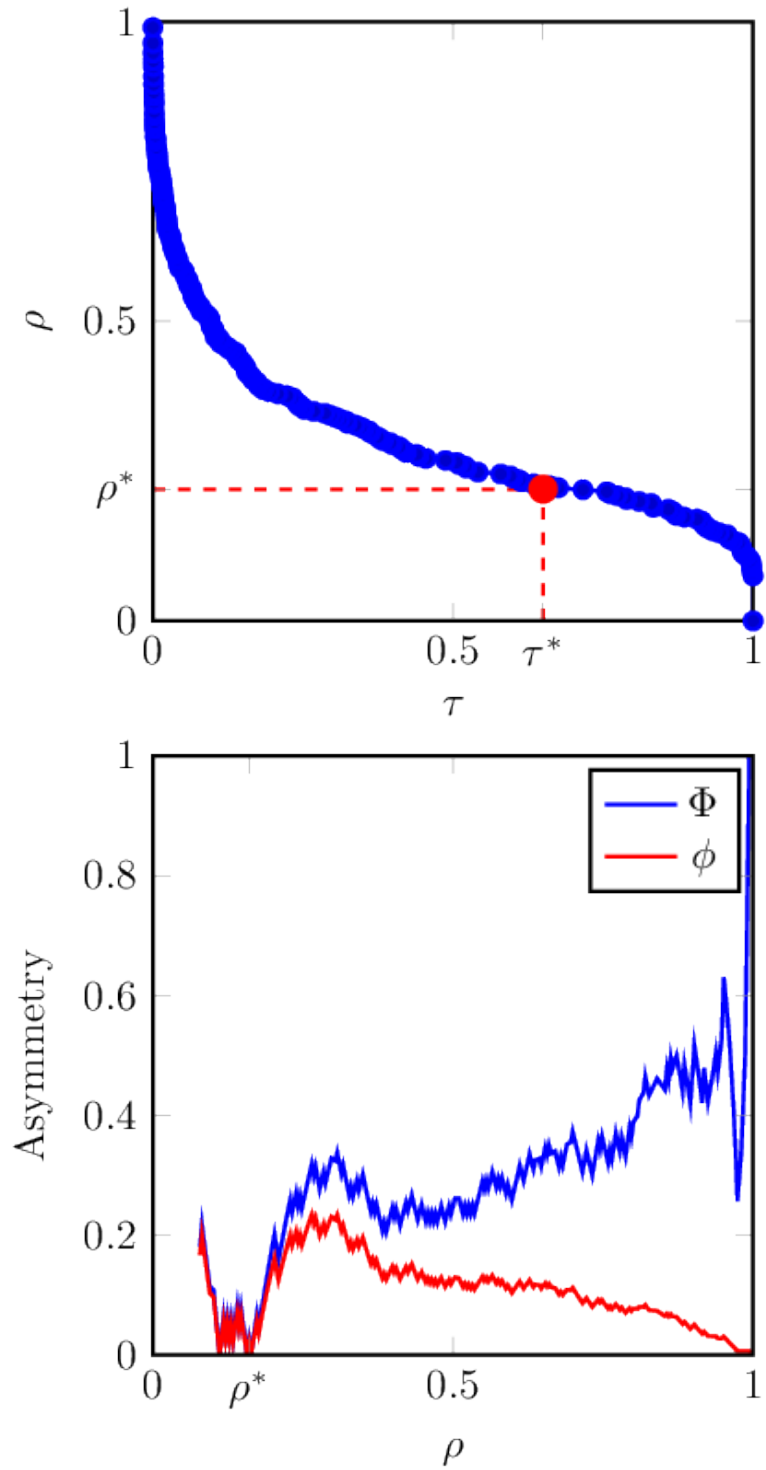

Fig.2

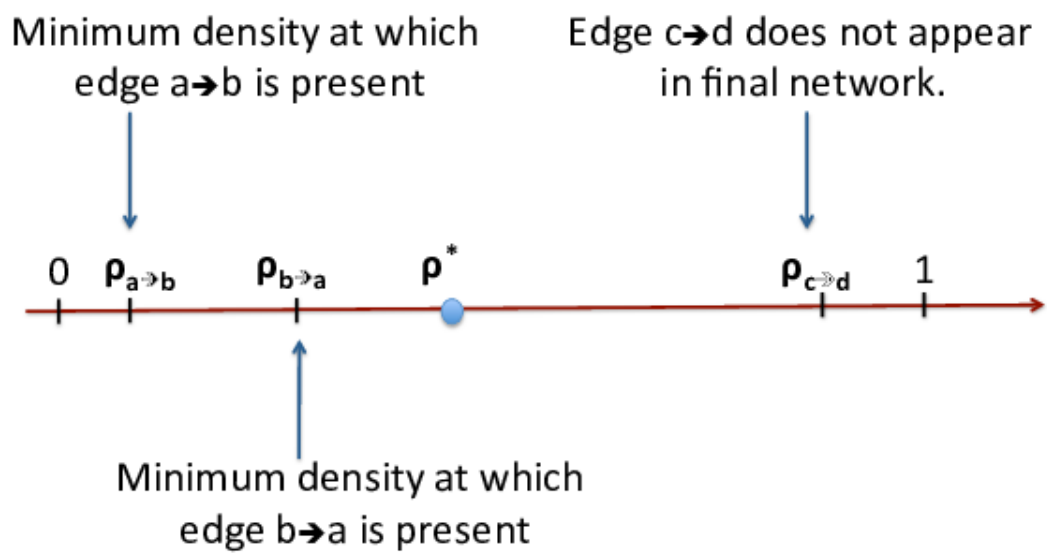

Fig.3

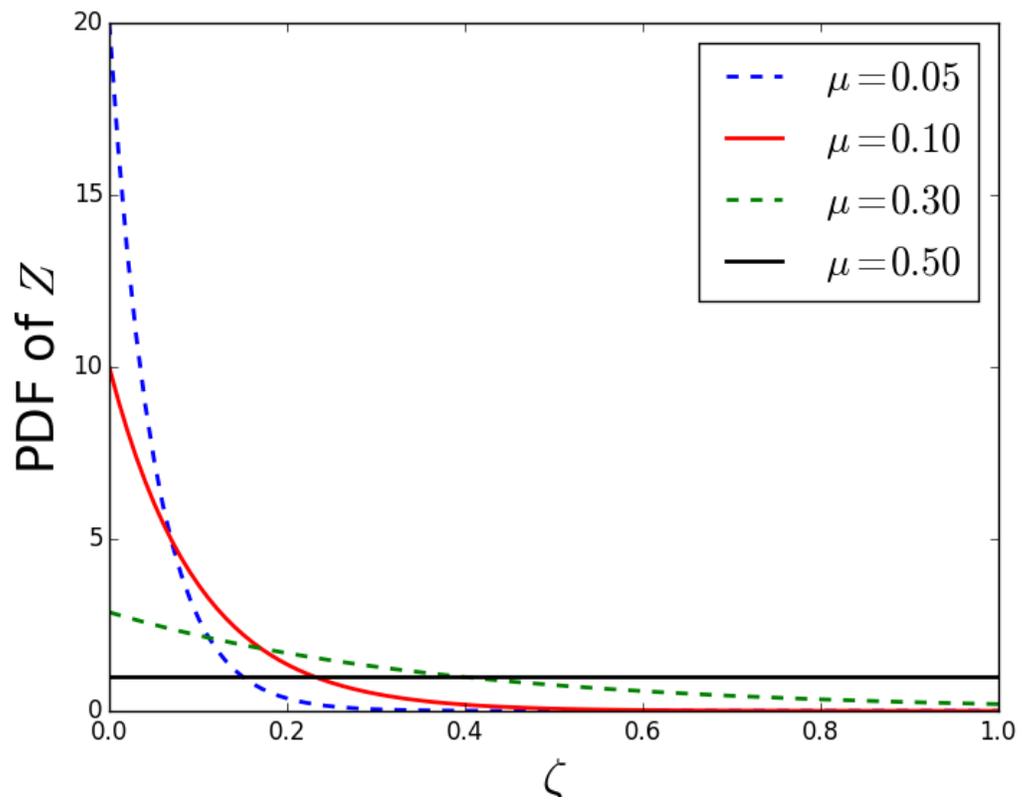

Fig.4

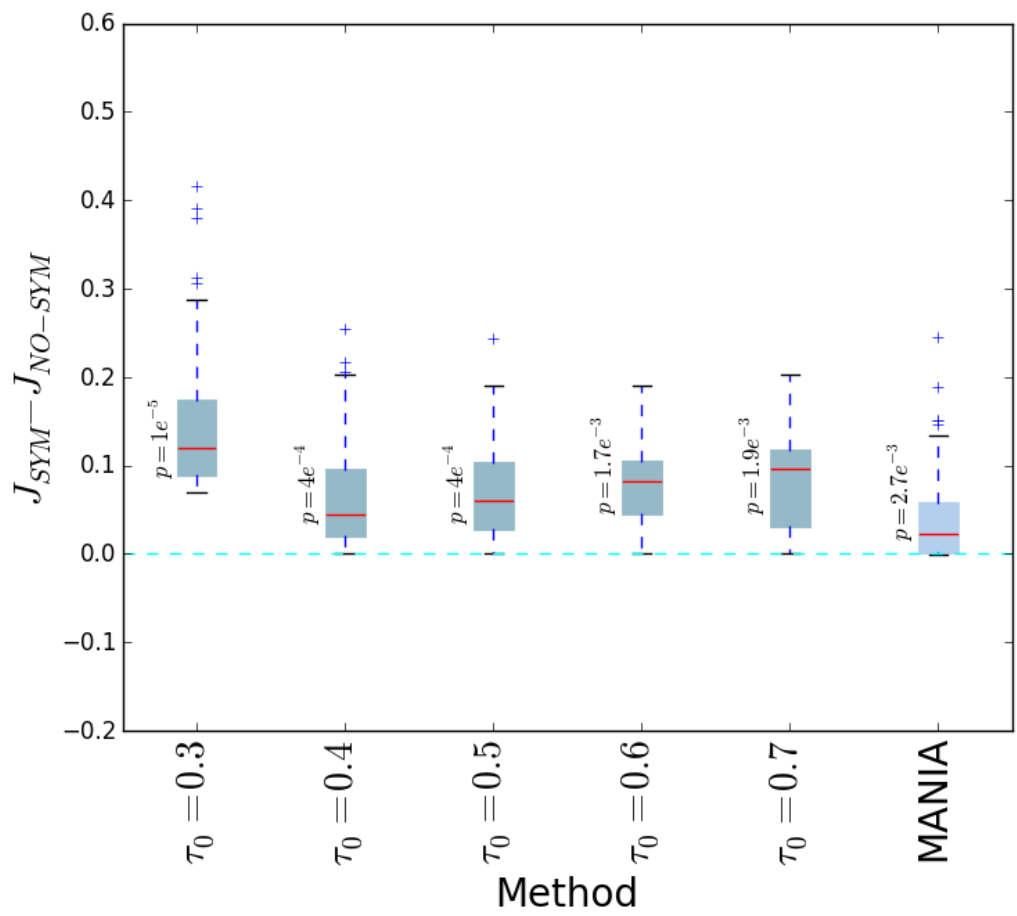

Fig.5

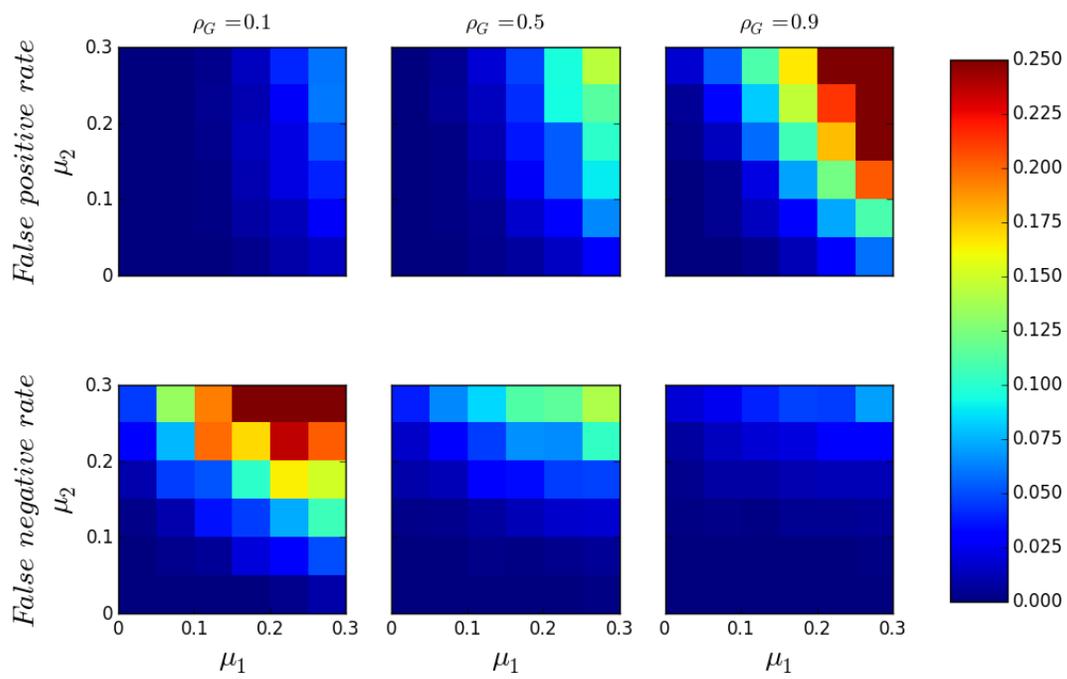

Fig.6

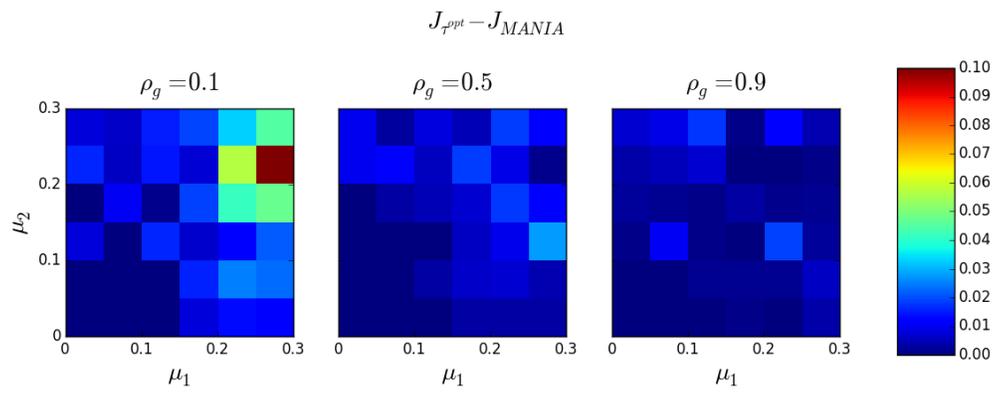

Fig.7

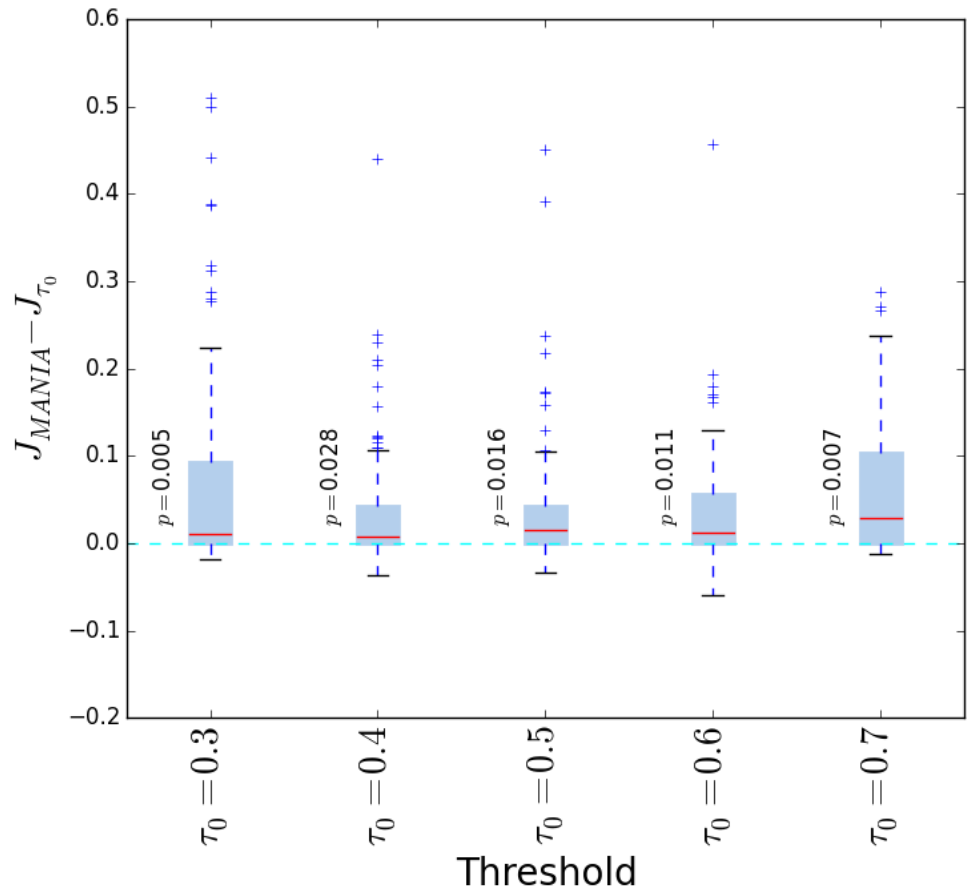

Fig.8

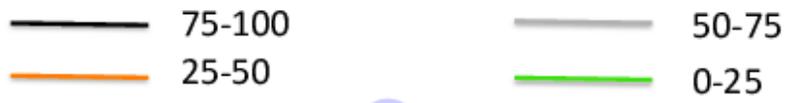

Percentage of subjects having the connection:
— 75-100   — 50-75
— 25-50   — 0-25

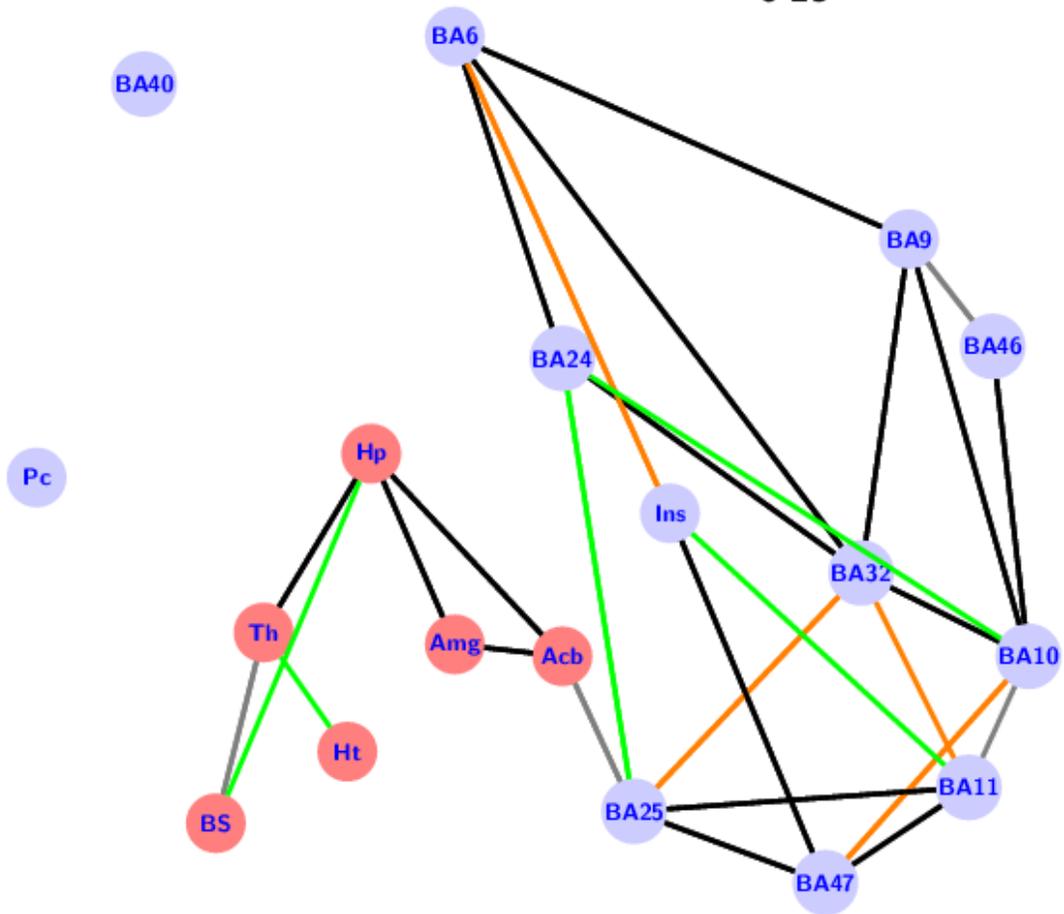

Fig.9

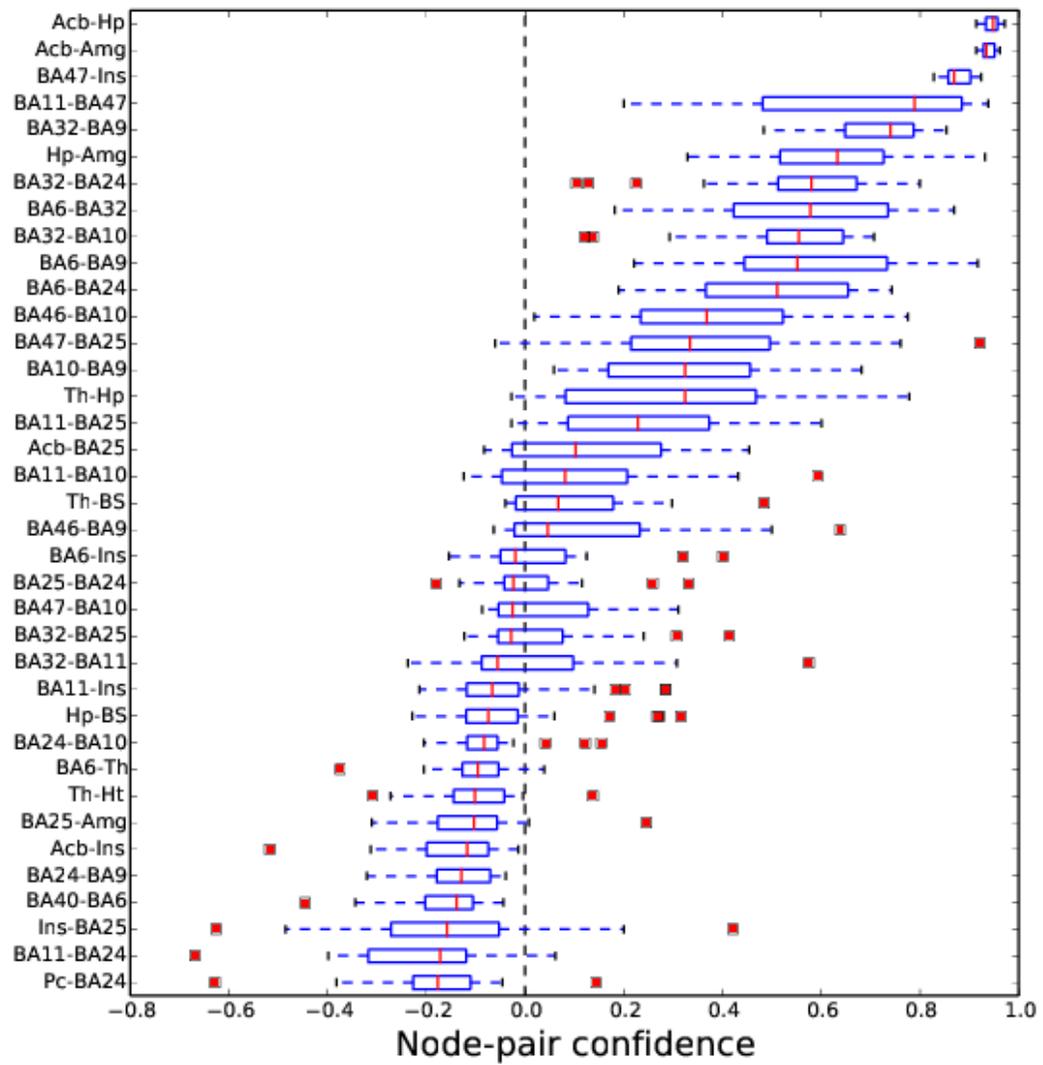

Fig.10

Table 1: The 18 corticolimbic ROIs we consider in the case-study.

| ROIs | Acronym | Number of voxels |
|---|---|---|
| Premotor cortex | BA6 | 3131 |
| Insula | Ins | 1858 |
| Ventromedial prefrontal cortex | BA10 | 1784 |
| Inferior parietal cortex | BA40 | 1598 |
| Dorsolateral prefrontal cortex | BA9 | 1422 |
| Mid-brain and pons | BS | 1406 |
| Orbito-frontal cortex | BA11 | 1243 |
| Thalamus | Th | 1100 |
| Hippocampus | Hp | 932 |
| Precuneus | Pc | 861 |
| Inferior prefrontal gyrus | BA47 | 851 |
| Ventral anterior cingulate | BA32 | 721 |
| Dorsal anterior cingulate cortex | BA24 | 593 |
| Dorsolateral prefrontal cortex | BA46 | 574 |
| Amygdala | Amg | 220 |
| Subcallosal cingulate | BA25 | 204 |
| Nucleus accumbens | Acb | 140 |
| Hypothalamus | Ht | 13 |

Table 2: Top-three nodes in rank-aggregated network based on four node-centrality metrics

| Centrality | Top-three nodes | | |
|---|---|---|---|
| Degree | BA32 | BA10 | BA25 |
| Closeness | BA25 | BA32 | Acb |
| Betweenness | BA25 | Acb | Hp |
| PageRank | BA10 | BA32 | BA11 |